\documentclass[aps,prl,amsmath,amssymb,reprint,superscriptaddress,floatfix]{revtex4-2}
\usepackage{graphicx}
\usepackage{dcolumn}
\usepackage{bm}
\usepackage{cancel}
\usepackage{silence}
\usepackage[hypertexnames=false]{hyperref}
\usepackage{xcolor}
\usepackage{empheq}
\usepackage{changepage}
\usepackage[normalem]{ulem}
\usepackage{amsmath}

\newcommand{\VM}[1]{\textcolor{blue}{#1}}
\newcommand{\RG}[1]{\textcolor{red}{#1}}

\newcommand{\NEW}[1]{\textcolor{orange}{#1}}

\newcommand{\UNI}[1]{#1}
\definecolor{brown}{rgb}{0.55,0.27,0.07}
\newcommand{\REV}[1]{#1}
\newcommand{\PUR}[1]{{\color{purple}#1}}

\newcommand{\edit}[1]{\textcolor{red}{#1}}        
\newcommand{\cut}[1]{\textcolor{red}{\sout{#1}}}  

\begin{document}

\preprint{APS/123-QED}

\title{Hydrodynamic Brachistochrone: Conflicting Paths of Time and Energy Minima within Viscous Media}
\author{Ramin Gasimli}
\affiliation{Department of Physics, University of Massachusetts, Amherst, MA 01003, USA.}
\author{Lei Yi}
\affiliation{Department of Physics, University of Massachusetts, Amherst, MA 01003, USA.}
\author{Shrabin Bajracharya}
\affiliation{Department of Physics, University of Massachusetts, Amherst, MA 01003, USA.}
\author{Anupam Pandey}
\affiliation{Department of Mechanical \& Aerospace Engineering, Syracuse University, NY 13244, USA.}
\affiliation{Department of Physics, Syracuse University, Syracuse, NY 13244, USA.}
\author{Varghese Mathai}  \thanks{vmathai@umass.edu}
\affiliation{Department of Physics, University of Massachusetts, Amherst, MA 01003, USA.}

\date{\today}

\begin{abstract}

We experimentally and theoretically study the
hydrodynamic analog of the classical brachistochrone problem: the {\it time-} and {\it energy-minimizing} paths for a
spherical particle rolling down an incline within a viscous fluid. We show that in the presence of viscous dissipation,
the paths of minima diverge from the classical cycloid, into curves of opposing curvature for time and energy, and are characterized by an effective dimensionless parameter, $St_p$, representing the ratio of the particle's viscous response time scale to its gravitational time scale.
Using a generalized variational framework, we show that the fastest path reduces to {nearly
straight ramps}, however, beginning and terminating in localized cycloids of curvature, $\kappa_c \sim St_p^{-2}$. 
Remarkably, the path of fastest descent on a given energy
budget requires navigating a non-monotonic path ({\it``S-shaped''}) with an interior point of inflection. Our findings reveal a unification of temporal and energetic optimality for transport through dissipative media, and expand the celebrated brachistochrone solutions to
the hydrodynamic regime.
\end{abstract}

\maketitle
\newpage

The brachistochrone problem, of determining the curve of fastest descent under gravity, stands as one of the foundational problems in classical mechanics. The earliest known posing of the problem is
attributed to Galileo Galilei, who in 1638 conjectured that the arc of a circle was
the curve of quickest descent. In 1696, Johann Bernoulli put forth the
challenge afresh~\cite{Bernoulli1696}, attracting interest from Newton, Leibniz, and
l'H\^{o}pital~\cite{Miguel1993}, which later played a crucial role in catalyzing the
development of the calculus of variations~\cite{Forsyth2012}.
Bernoulli's celebrated solution, the ``cycloid'', has proven to be remarkably robust, remaining the
shortest-time path in a variety of contexts and even in the presence of several additional constraints including
inertial rolling, Coulomb friction, and non-uniform gravitational
fields~\cite{Pessoa2024,Legeza2010,haws1995exploring,Ashby1975,Hayen2005,Lipp1997,Salinic2009,Paladi2023}. 

A few studies have also explored the brachistochrone in settings where fluid--structure
interactions exist, such as a fluid-filled cylinder
rolling~\cite{Balmforth2007,Gurram2019} in inviscid media or on flexible substrates~\cite{Camassa2008,Aristoff2009}, or in settings where fluid drag plays a
role~\cite{VratanarSaje1998,Avron2004,Tam2007,Quinn2015}.  {\color{black}For the case of rolling within viscous media, a companion minimization problem emerges, that of the {\it path of lowest energy dissipation}, which can be of broad relevance in self-propulsion and in transport processes~\cite{Mandre2020,Mandre2022}. {The time-versus-energy question arises in a variety of settings,} and the coupling with energy minimization can guide design rules
for efficient transport of particles, droplets, and bubbles through viscous
fluids~\cite{Mertaniemi2011, bouillant2018leidenfrost, bico1999pearl, mahadevan1999rolling,Shankar2022, Sobolev2023,mathai2020bubbly}, with applications ranging from targeted drug delivery to
lab-on-a-chip systems.} In such contexts, performance depends critically on steering particles to
a target with minimal energy loss and often in the shortest time. Despite its broad relevance, the question of whether viscous
media give rise to {\it distinct} paths that separately minimize travel time and
energy loss has remained unexplored.

In this Letter, we present experiments and theory for the minimal-time and minimal-energy-loss
 paths for a spherical particle rolling under gravity within a viscous fluid. {We show that viscous
dissipation fundamentally breaks the classical brachistochrone solutions: the shortest-time paths deviate from cycloids and evolve toward composite curves that combine linear ramps with localized cycloidal end-caps. In contrast, the paths that minimize total energy loss are of
opposite curvature (inverted cycloids) to those of time-minima. Strikingly, when both time and energy constraints coexist, we reveal a family of intermediary paths of rolling, with local inflection points (``S-shaped'' tracks).}

\begin{figure}[!t]
\centering{\includegraphics[width=0.48\textwidth]{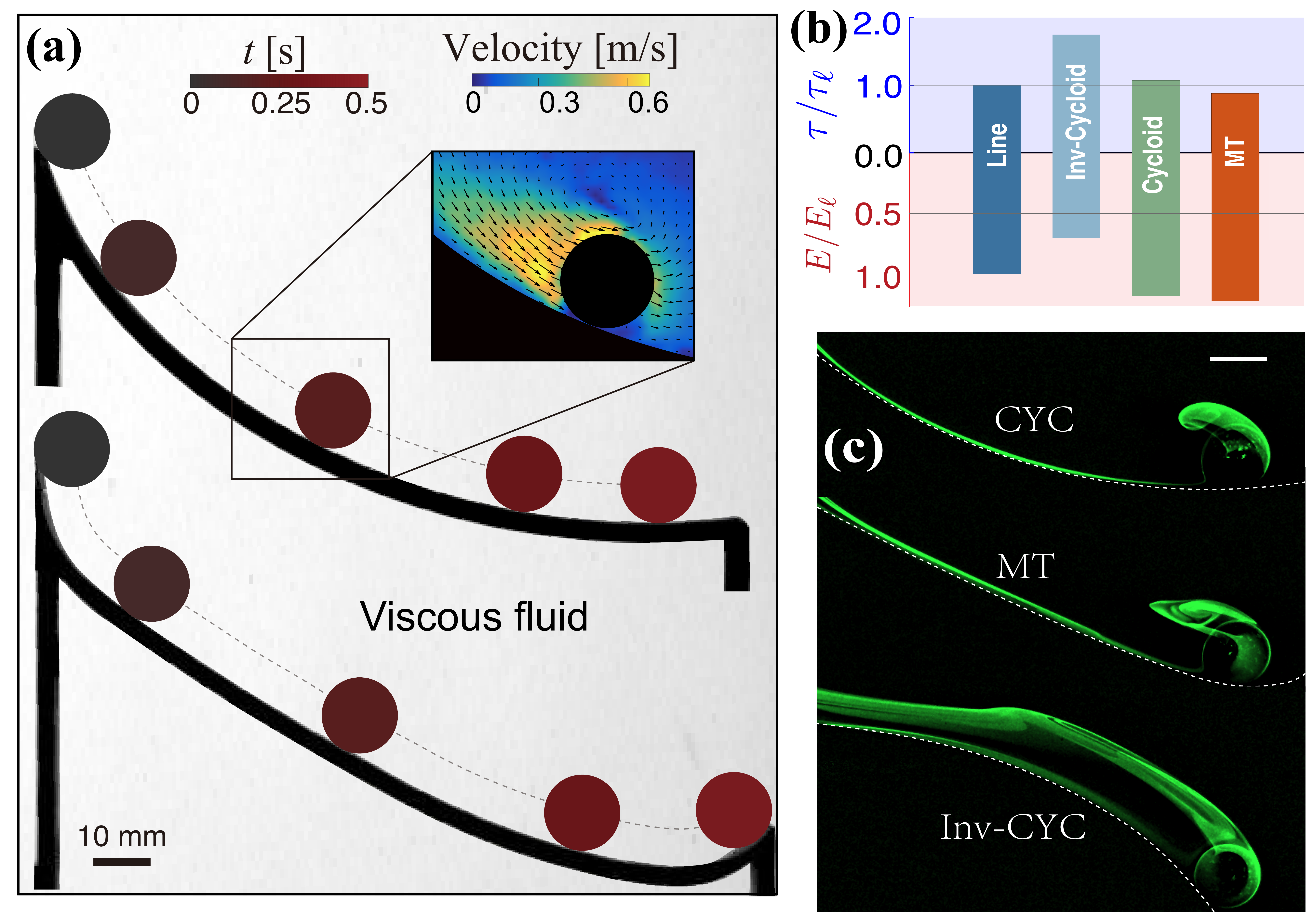}}
\caption{\label{fig01_nature_ideal}
Experiments on paths of shortest time and minimal energy loss in a viscous fluid.
(a) Time-lapse composite: steel ball on cycloid ramp (upper) and on MT ramp (lower). Inset: Snapshot of velocity field around the sphere, obtained from particle image velocimetry (PIV). The MT ramp
is 24\% faster, despite its 10\% longer path length. (b) Comparison of normalized descent time (+ve y-axis) and normalized dissipated energy (-ve y-axis) for various ramps. The cycloid is neither the fastest nor the least energy-consuming path.
(c) Fluorescent dye visualization of the wake vortex for the {three ramp shapes (cycloid, MT, and inverted cycloid)},
revealing complex wake structures left behind by the path of the rolling sphere.
See also {Movies S1 and S2}~\cite{supp}. Scale-bar: $10~$mm.}
\end{figure}

Figure~\ref{fig01_nature_ideal} shows experimental measurements of a steel sphere rolling down an incline within a viscous liquid (70-30 volume ratio of glycerol-water mixture). A time-lapse
composite of the sphere descending along two different 3D-printed tracks is shown in Fig.~\ref{fig01_nature_ideal}a:
the upper track is the ``classical cycloid'', and the lower one is a ramp (MT) that is flatter in the middle, but more sharply curved at the start- and end-points. Notably, the MT ramp beats the cycloid by a 24\% margin in descent time despite the curve's length being 10\% greater than that of the cycloid.
In {Fig.~\ref{fig01_nature_ideal}b we compare} the travel time and energy dissipated for four different tracks, namely a line ramp, a {``cycloid''}, an ``inverted cycloid'', and the MT ramp. A striking conflict emerges between travel time and
energy loss: the MT ramp is the quickest, but also dissipates the most
energy, while the most energetically economical ramps are the slowest. The cycloid is neither the fastest nor the least
dissipative (see Movie S1). This conflict is not apparent a priori, since both the travel time $T = \int ds/v$ and energy loss $E_{\text{loss}}= \int F\, ds$ depend on the full path. A theoretical prediction for inviscid rolling at the given mean slope in the experiments ($m = 0.577$ as in Fig.~\ref{fig01_nature_ideal}a,b,c), corresponding to a mean angle with the horizontal of $\theta_m = 30^\circ$, suggests that the cycloid ought to be significantly faster (Fig.~\ref{Fig_Supp_inviscid_length_time_ratios}
\cite{supp}). Hence, the relative deviation in the ratio of times is around 73\% compared to the inviscid rolling case.  Accounting for {the} particle's buoyancy, inertia,
or non-zero initial release velocity ($v_i \neq 0$) cannot explain these differences; hence these ought to have a hydrodynamic origin.

\vspace{0.1cm}
 Figure~\ref{fig01_nature_ideal}c shows the accumulated wake structures for three of these paths: the cycloid, the inverted cycloid, and the MT track. The far wake patterns are notably distinct for the three ramps (see Movie S2). However, the near-field wake structures are similar across the ramps. The wake flow behind the sphere, as captured by particle image velocimetry (PIV){,} is shown in Fig.~\ref{fig01_nature_ideal}a, inset{,} {suggesting} the role of the hydrodynamic effects during viscous rolling.
\begin{figure}[!b]
\centering{\includegraphics[width=0.49\textwidth]{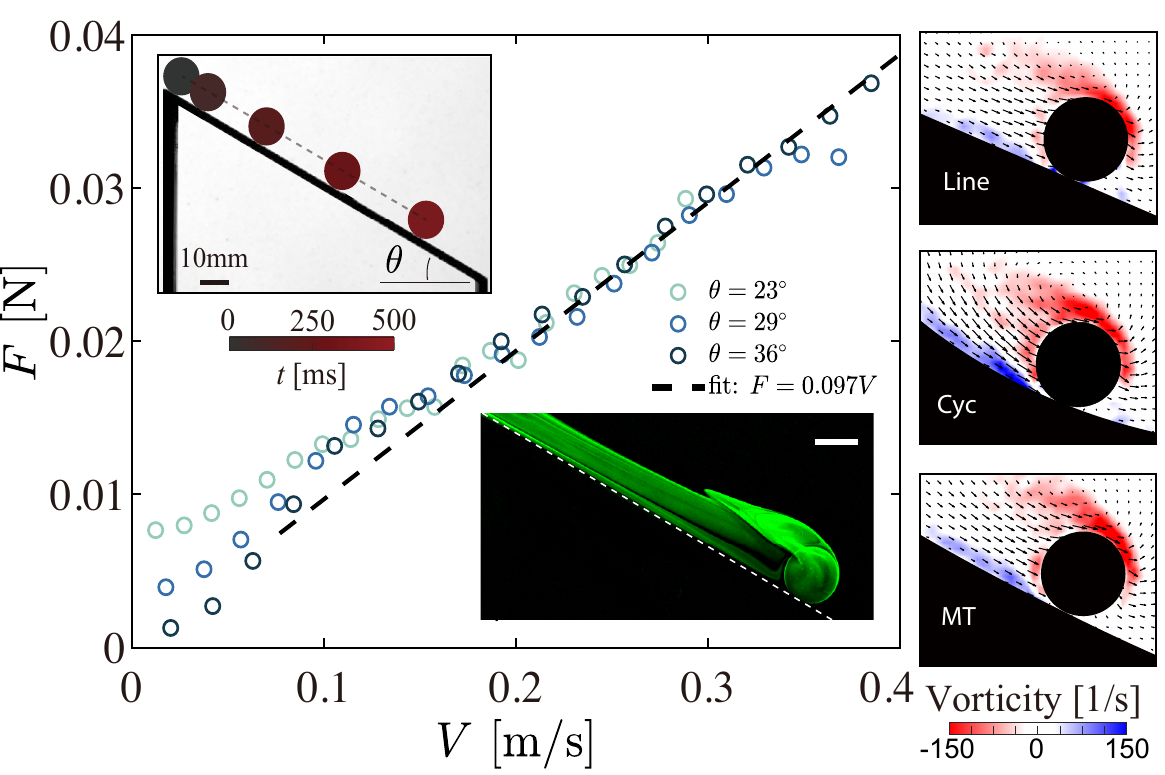}}
\caption{Experimentally estimated drag vs.\ velocity on ramps of different slopes.
Insets: (upper left) Snapshots of a rolling sphere along a line ramp, separated by 0.125 s time intervals and (lower right)
wake pattern left behind by fluorescent dye ($10~$mm for the {scale} bar; see also Movie S3). The right side images show PIV snapshots suggesting a very similar wake-vortex structure for line, cycloid, and MT ramps.}
\label{drag_exp}
\end{figure} 
\begin{figure*}[!ht]
\centering{\includegraphics[width=1\textwidth]{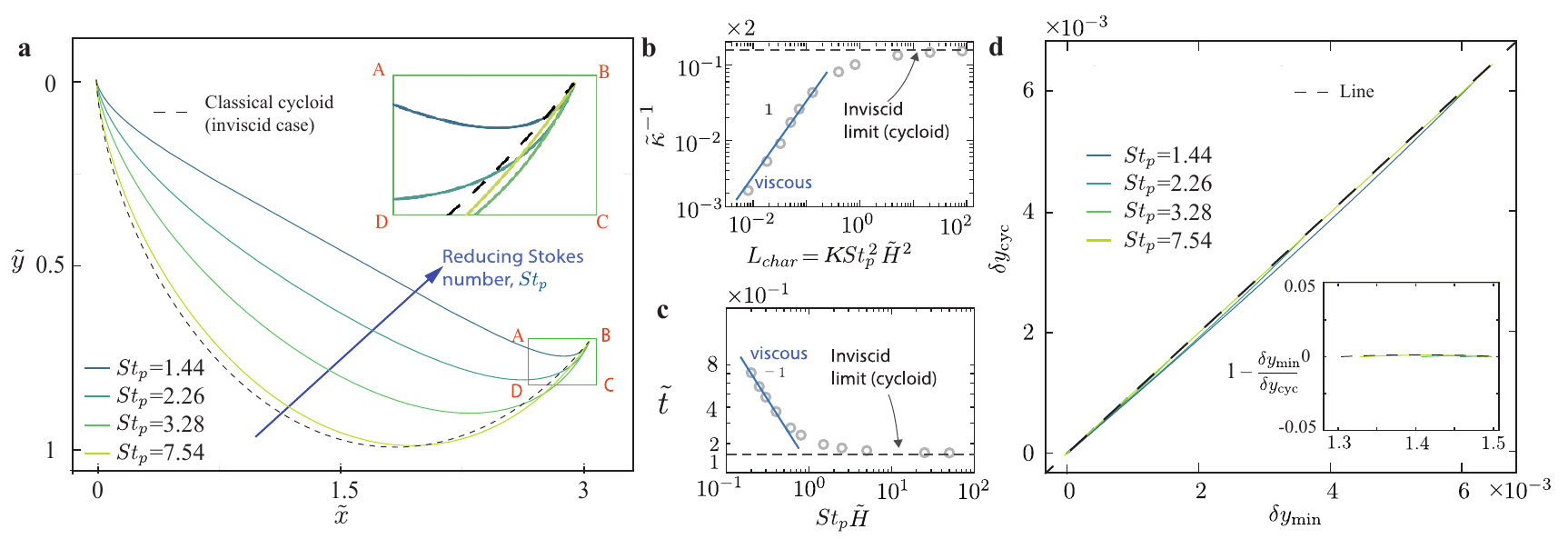}}
\caption{Theoretically predicted optimal ramps (Eq.~\ref{eq:kappa_mu}) in a viscous fluid for
different Stokes numbers $St_p$.
(a) Time-minimizing ($\Pi=0$) ramps: as $St_p$ decreases, the optimal ramp shifts from a
cycloid toward line-like ramps with sharp end curvature (inset A--B--C--D).
(b) Local curvature scaling $1/\kappa$ vs.~characteristic length $L_{char}=St_p^2\tilde{H}^2$. {The sloped solid line is the viscous prediction $1/|\kappa_f| = K\,St_p^2\tilde{H}^2$ with $K=2\tan\theta_f$, and the dashed horizontal line marks the inviscid (cycloid) plateau.}
(c) Non-dimensional travel time $\tilde{T}$ scales as $L_{char}^{-1/2}$ in the viscous
limit.
(d) Deflection $\delta_{y_{\min}}$ of the optimal ramp from a straight line, plotted against
deflection $\delta_{y_{cyc}}$ of the best-fit local cycloid of curvature $1/L_{char}$. Data
falling on the 45$^\circ$ line confirms the end-caps are locally cycloidal; deviations in the
compensated inset quantify the mid-section departure.}
\label{fig_timemain}
\end{figure*}
We express these through a dynamical equation of motion for the particle that includes gravity, buoyancy, fluid added mass, translational
and rotational inertia, and hydrodynamic drag, as:
\vspace{-0.2cm}
\begin{equation}
    \Big(m+m_{a}+{\tfrac{I}{r_p^2}}\Big)\frac{dv}{dt}=(m-\rho_f \mathcal{V})g\sin\theta-F(v),
\label{eq:EoM}
\end{equation}
where $F(v)$ is the drag force, $m$ and $r_p$ are the particle's mass and {radius} respectively, $I$ is its rotational inertia,
$\rho_f$ is the density of the fluid, $\mathcal{V}$ is the volume of the object, $g$ is the
gravitational acceleration, $v$ its instantaneous velocity, and $m_{a}$ the added mass. Figure~\ref{drag_exp}a shows the estimated drag versus velocity for ramps of different inclinations. The data approximates well onto a single, nearly linear relation, $F \approx 0.094\,\eta\,r_p v$. {Linear drag law is a good approximation for the low particle Reynolds numbers ($\mathrm{Re}_p \sim 0.5$--$2$) in our experiments, for which inertial corrections to the Stokes drag remain small and {history effects are negligible \cite{mathai2020bubbly}}. Due to sphere rotation and the squeeze-film flow in the thin gap as it rolls against the boundary, the viscous drag prefactor is $\approx 2.5 \times $ that of Stokes drag of a sphere in unbounded domain}~\cite{GoldmanCoxBrenner1967}. The PIV-measured
vorticity fields (Fig.~\ref{drag_exp}, right insets and Movie S4) reveal strongly squeezed viscous flow, with a wake
structure that is similar across the different ramp shapes. {Since the drag force can be expressed as $F\propto v$, it permits building a generalized variational framework as follows.} Non-dimensionalizing (with $\tilde{}$ variables) using $v=\sqrt{2gL}\tilde{v}$,
$s=L\tilde{s}$, $t=\sqrt{{L}/{2g}}\,\tilde{t}$ gives
$d\tilde{v}/d\tilde{t}=B\sin\theta-A\tilde{v}$.
Here, $B=\tfrac{\Gamma-1}{1+\tfrac{14}{5}\Gamma}$, and the dimensionless drag coefficient
$A$ follows from the experimentally measured linear relation in Fig.~\ref{drag_exp}.
The drag coefficient
is $\tilde{F}=\tfrac{18\,c_f\sqrt{\Gamma-1}}{\sqrt{\tilde{r}}\,\mathcal{G}a(1+\tfrac{14}{5}\Gamma)}\tilde{v}$,
with $c_f=2.1$. Here $\Gamma=\rho_p/\rho_f$ is the particle to fluid density ratio, and $\mathcal{G}a=\sqrt{\frac{8(\Gamma-1)r_p^3g}{\nu^2}}$ is the Galileo number, representing the ratio of gravitational forces to viscous forces.

In most practical settings, the constraints on time and energy loss may coexist, as in situations desiring
arrival at a target destination quickly while not expending much energy, e.g. steering a bubble through a microfluidic device~\cite{zhang2024boosting}. {Because} the drag law is nearly linear in $v$, the dissipated power $\dot{E} \propto v^2$ can be treated as a viscous potential. Therefore,  we can propose a generalized variational framework for the curves of
shortest time and least energy loss.  This yields a combined
functional for time and energy, given by
\begin{multline}
\mathcal{I}_\mu=\int_0^{\tilde{T}} \Big[ 1 + \mu A\tilde{v}^2 + \lambda
\Big(\frac{d\tilde{v}}{d\tilde{t}} + A\tilde{v} - B\sin\theta\Big) \\
+ \sigma_x \Big(\frac{d\tilde{x}}{d\tilde{t}} - \tilde{v}\cos\theta\Big)
+ \sigma_y \Big(\frac{d\tilde{y}}{d\tilde{t}} - \tilde{v}\sin\theta\Big) \Big]\,\mathrm{d}\tilde{t},
\label{eq:Imu}
\end{multline}
where $\int_0^{\tilde{T}} 1 \ {\mathrm{d}\tilde{t}}$ minimizes time of travel under a set of kinematic constraints, dynamic constraints, and energy constraints using Lagrange multipliers, $\sigma_{i}$, $\lambda$, and $\mu$, respectively (see also supplemental~\cite{supp}). The first variation of $\mathcal{I}_\mu $ yields the
path curvature, $\kappa = d \theta/ ds$, as the central geometric quantity,
\begin{equation}
\kappa = \underbrace{-\frac{B\cos\theta}{\tilde{v}^2} +
\frac{2A\cos\theta\sin(\theta-\theta_f)}{\tilde{v}\cos\theta_f}}_{\kappa^{(T)}} +
\Pi\cdot\underbrace{\frac{2B\cos^2\theta}{\tilde{v}\,\tilde{v}_f\cos\theta_f}}_{\kappa^{(E)}},
\label{eq:kappa_mu}
\end{equation}
where
$\Pi \equiv \mu A\tilde{v}_f^2/(1+\mu A\tilde{v}_f^2) \in [0,1)$
runs from $\Pi=0$ (quickest descent) to $\Pi\to 1$ (energy-minimizing path).
The first two terms, {$\kappa^{(T)}$,} represent the negative curvature of the pure brachistochrone (quickest descent): a downward dip as expected for the classical inviscid problem ($A \to 0$ limit); while the third term due to an energy constraint adds a positive curvature, since $\kappa^{(E)}>\,0$.

\noindent
{\it Paths of quickest descent:}
 In the curvature equation (Eq.~\ref{eq:kappa_mu}), setting $\Pi=0$ recovers the global time-minimum paths for a variety of viscous dominated regimes. A dimensionless parameter, $St_p =  (B/(A^2 \tilde{H}))^{1/2}$, representing the ratio of the viscous time scale of {the} particle to its gravitational time scale, emerges as the single control parameter in the problem. The optimal path, from Eq.~\ref{eq:kappa_mu}, is therefore a composite: cycloidal end-caps where inertia dominates,
stitched to a nearly linear mid-section where the viscous effects dominate. {Figure~\ref{fig_timemain}a shows a family of optimal paths for varying $St_p$, illustrating} the gradual departure from the classical cycloid as viscous effects become important (or {decreasing} $St_p$). The MT track in our experiments (Fig.~\ref{fig01_nature_ideal}a) which outperformed the cycloid by 24\% represents an example of this optimal travel time. The distinctively sharp curvatures at the start and
endpoints are visible for the MT track, joined by a comparatively straighter mid-section.
Toward the endpoint, the viscous contribution vanishes and
$\kappa_f^{(T)} = -B\cos\theta_f/\tilde{v}_f^2 < 0$ renders the end-cap curvature sharply concave (negative).

Dimensional considerations in the viscous regime ($St_p \ll 1$) suggest that the characteristic length scale of particle acceleration should be relatable to the curvature $\kappa_f$ of the optimal path. This yields a prediction $\kappa^{-1} \propto St_p^2$, which holds well in the viscous regime and until the curvature plateaus to that of
the inviscid cycloid at large $St_p$ (Fig.~\ref{fig_timemain}b). Similarly, travel time scales
as $\tilde{T}\sim L_{char}^{-1/2}$ in the  viscous limit (Fig.~\ref{fig_timemain}c). Since the endpoints are inertia-dominated, we further show that {the} {end-caps} of the path are locally cycloids. In Fig.~\ref{fig_timemain}d, the axes
compare deflection of the optimal curve from a local cycloid of
curvature $1/L_{char}$, with data on the 45$^\circ$ line confirming a local-cycloid
shape.

\begin{figure}[!t]
\centering{\includegraphics[width=0.47\textwidth]{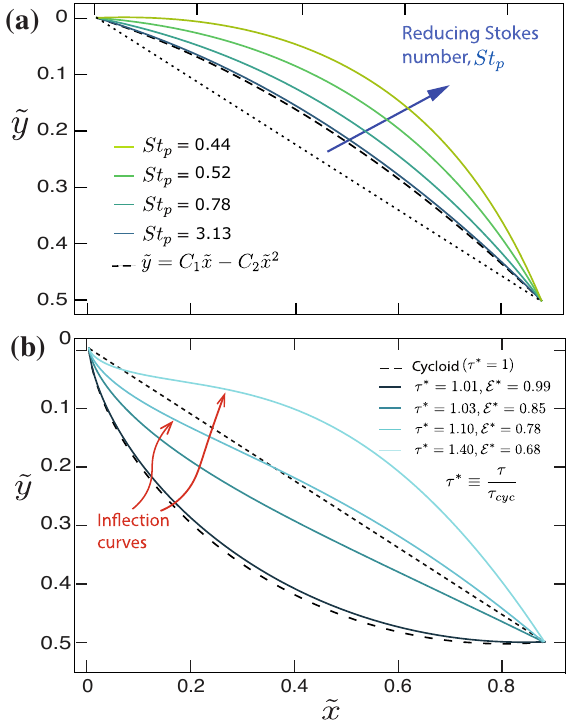}}
\caption{\label{fig03_nature_ideal}
Energy-minimizing and energy-constrained ramps. (a) Energy minima ($\Pi \to 1$) for
varying particle Stokes number yield inverted cycloid-type ramps, consistent with experiments
(Fig.~\ref{fig01_nature_ideal}b). The curvature follows from the $\Pi\to 1$ limit of
Eq.~(\ref{eq:kappa_mu}).
(b) Energy-constrained time minima ($0<\Pi<1$) in the inviscid limit ($A\to 0$):
the classical cycloid (dashed) evolves through a family of paths with increasing $\Pi$,
developing a local inflection (``S-shaped'' paths) for $\Pi>1/2$. In the $A\to 0$ limit, the
energy-minimizing trajectory reduces to a parabola (see supplemental
Fig.~\ref{Fig_parabola_supp}~\cite{supp}).}
\end{figure}

\noindent
{\it Paths of least energy-loss:} {Having mapped the time-optimal paths, we now turn to the opposite objective, where energy expenditure rather than travel time is minimized. The same variational framework applies, but the shape of the optimum inverts: the concave cycloidal end-caps of the fastest path give way to convex, inverted-cycloid end-caps.}
The experiments showed that a convex-upward, arch-shaped profile yielded the lowest energy dissipation among the four tested tracks in Fig.~\ref{fig01_nature_ideal}b,c, bar
graph. {Setting} $\Pi\to 1$
in Eq.~\ref{eq:kappa_mu} gives an analytical prediction for the curvature of the energy-minimizing path.
The $\kappa^{(E)}$ term is dominant
(see supplemental Eq.~{S-18}~\cite{supp}). Since $\kappa^{(E)}>0$ everywhere along the
path, the energy-minimizing path is {strictly more} convex than the time minimizing (fastest) path. The physical consequence of this is sharpest at the endpoint where the viscous contributions vanish (as $\theta \to \theta_f$), giving
\begin{align}
\kappa^{(E)}_f = +\frac{B\cos\theta_f}{\tilde{v}_f^2} > 0, \qquad
\kappa^{(T)}_f = -\frac{B\cos\theta_f}{\tilde{v}_f^2} < 0,
\label{eq:kappa_energy_sign}
\end{align}
{such that} $\kappa^{(E)}_f = -\kappa^{(T)}_f$, i.e. the cycloidal end-cap of the classical brachistochrone is replaced by an inverted cycloid {end-cap} for the energy-minima paths. These rationalize how having an energy-saving goal (Fig.~\ref{fig03_nature_ideal}a) gives rise to strikingly different paths from those of time minima, consistent with the experimentally observed conflict in
Fig.~\ref{fig01_nature_ideal}b.
With decreasing $St_p$, the ramps become increasingly curved and farther from the time minima paths.
Interestingly, in the limit of {\it least-dissipation} ($St_p\gg 1$, $A\to 0$), the energy-minimizing
trajectory from Eq.~\ref{eq:kappa_mu} reduces to the familiar parabolic path of projectile motion
(see Fig.~\ref{Fig_parabola_supp}~\cite{supp}). 

\vspace{0.1 cm}

\noindent
{\it Quickest descent under an energy constraint:}
It is deduced from Eq.~\ref{eq:kappa_mu} that as the energy budget tightens (with increasing $\Pi$), the end-cap curvature
$\kappa_f=(B\cos\theta_f/\tilde{v}_f^2)(2\Pi-1)$ undergoes a sign reversal at a critical value of $\Pi=1/2$. However, near the start of the ramp the particle is still gaining in speed (since ${\tilde{v}_i} \to 0$), and hence the inertial
curvature $\kappa^{(T)} \sim B\cos\theta/\tilde{v}^2$ dominates, with negative curvature (concave) at the start of the rolling. For
$\Pi>1/2$ the path must begin concave and finish convex: i.e. it must undergo a reversal of curvature, or an inflection point along the path ({\it ``S-shaped''}). {The inflection point appears at $\Pi^*=1/2$, where it sits at the endpoint of the path and  migrates inward as $\Pi$ increases beyond $1/2$.} These
{\it S-shaped} paths are, therefore, a distinctive manifestation of the conflicting effects of time (concave) and energy (convex) minima.
Figure~\ref{fig03_nature_ideal}b illustrates a family of {paths} as solutions, where the transition from a {\it cycloid-like} path to an {\it S-like} path (a path containing an interior inflection point) emerges once the energy constraint becomes increasingly more stringent. The legend indicates the normalized time of travel and energy loss for each path.

\vspace{-0.1cm}
In summary, our experiments and theoretical framework have revealed that viscous hydrodynamics fundamentally
alters the paths of optimal descent for a particle in a fluid. We show experimentally that cycloids can be
outperformed by up to 24\% in the time of descent, with time and energy
optima paths in direct conflict (Fig.~\ref{fig01_nature_ideal}a,b). The paths of quickest descent and those of least energy-loss have equal and opposite end-point curvature. The family of solutions (Fig.~\ref{fig_timemain}) collapses under a single control parameter, $St_p$, representing the ratio of the particle's viscous
response time scale to gravitational free fall time.
Strikingly, we show that paths optimized for time, but under a finite energy budget, e.g. for a fuel-limited transport problem, give rise to an ``S-shaped'' navigation strategy, where the particle must
first descend steeply and then arch upward before curving down toward its endpoint (Fig.~\ref{fig03_nature_ideal}b). These arise as a direct
consequence of the time--energy conflict of hydrodynamic transport. 

{The unified variational framework introduced here, and in particular the single-parameter interpolation between time and energy optima, suggests natural extensions to several open problems in viscous transport. For active microswimmers navigating viscous environments under a finite metabolic budget, the S-shaped paths identified here may provide a physical basis for understanding non-trivial swimming trajectories that balance speed against energy expenditure. More broadly, the competing curvatures of time- and energy-optimal paths could inform the design of microfluidic geometries for particle sorting and targeted delivery, where channel shape must be optimized against both throughput and driving power. Ongoing work addresses the rolling of two-dimensional droplets where similar regimes of time and energy minima have been observed. We note that the present framework can be extended to regimes of nonlinear drag, where $\tilde{F}\sim \tilde{v}^n$. However, the linear drag regime discussed here can already capture the essential ingredients for the {time--energy} conflict.}\vspace{0.05cm}

\noindent
We are grateful to Nick Wang and Emily Aslanian for initial developments of the experiment. We thank Patrick Jefferson, Micayah Ritchie, Sander Huisman, Kuntal Patel and Xiaojue Zhu for useful discussions. {RG acknowledges support from the Commonwealth Honors College (CHC). VM acknowledges support from NSF Award \#2340293.}

\bibliographystyle{apsrev4-2}
\bibliography{refs_v5}

\clearpage
\newpage
\setcounter{equation}{0}
\setcounter{section}{0}
\setcounter{figure}{0}
\setcounter{table}{0}
\setcounter{page}{1}
\makeatletter
\renewcommand{\theequation}{S-\arabic{equation}}
\renewcommand{\thefigure}{S-\arabic{figure}}
\renewcommand{\thesection}{S-\Roman{section}}
\renewcommand{\thesubsection}{S-\Roman{section}-\alph{subsection}}
\renewcommand{\thetable}{S-\arabic{table}}

\providecommand{\UNI}[1]{#1}
\providecommand{\NEW}[1]{#1}
\providecommand{\VM}[1]{\textcolor{blue}{#1}}
\providecommand{\RG}[1]{{\color{red}#1}}
\definecolor{brown}{rgb}{0.55,0.27,0.07}
\providecommand{\REV}[1]{#1}
\providecommand{\PUR}[1]{{\color{purple}#1}}
\providecommand{\edit}[1]{\textcolor{red}{#1}}
\providecommand{\cut}[1]{\textcolor{red}{\sout{#1}}}

\clearpage
\onecolumngrid
\begingroup

\begin{center}
{\large \textbf{Supplemental Material}}\\[3pt]
\end{center}

\begin{adjustwidth}{1.2cm}{1.2cm}
\fontsize{10}{14}\selectfont

\setcounter{equation}{0}
\setcounter{section}{0}
\setcounter{figure}{0}
\renewcommand{\theequation}{S-\arabic{equation}}
\renewcommand{\thefigure}{S-\arabic{figure}}
\renewcommand{\theHequation}{S.\arabic{equation}}
\renewcommand{\theHfigure}{S.\arabic{figure}}

\newcommand{\LY}[1]{\textcolor{blue}{#1}}

\vspace{0.5cm}
\noindent
{\bf Supplemental Videos}

{
\noindent{\it Movie S1.}
Comparison of the steel sphere rolling along the cycloid, the minimum-time (MT), the line, and the inv-cycloid ramps, demonstrating that $t_{MT}<t_{l}<t_{cyc}<t_{i\text{-}cyc}$. {(\texttt{Movie S1})} \href{https://www.dropbox.com/scl/fi/ru6pbh60p7v4pcztrfm32/Video-SM1.mp4?rlkey=ok9y4h5h5yw9ciqn13v3n8g98&dl=0}{\tt [link]}\\
\noindent{\it Movie S2.}
Videos showing fluorescent dye visualization of the wake vortex structures generated along the three ramps, cycloid, MT, inv-cycloid, shown in sequence. {(\texttt{Movie S2})}\href{https://www.dropbox.com/scl/fi/bct6t50tukhmmorjl7owc/Video-SM2.mp4?rlkey=pvhzoiodfyfcu4s4gbb83abgb&dl=0}{\tt [link]}\\\noindent{\it Movie S3.}
The video shows fluorescent dye visualization of the wake vortex structures generated along the line ramp. {(\texttt{Movie S3})}\href{https://www.dropbox.com/scl/fi/05am0fwjr6wys2dec91mk/Video-SM3.mp4?rlkey=9nmjd1o4163boumqbguer39hu&dl=0}{\tt [link]}\\
\noindent{\it Movie S4.}
PIV measurements of the vorticity field around the rolling sphere on
the line, the cycloid, and the MT ramps.
Despite variations in ramp geometry, the near-sphere wake
structure remains qualitatively similar, supporting the use of a universal linear drag law (Fig.~2 of the main text). {(\texttt{Movie S4})}\href{https://www.dropbox.com/scl/fi/91a1lls3obycb3hsv75zt/Video-SM4.mp4?rlkey=nm5uojsy0fyxr6mpfrbykjf4n&dl=0}{\tt [link]}
}

\vspace{0.5cm}
\noindent
{\bf Inviscid rolling:} For the cycloid shape, we have $y=R(1-\cos\phi)$ and $x=R(\phi-\sin\phi)$.
Total time $\int dt=\int ds/v$ is:
\[
t_{cyc} = \int\frac{\sqrt{(dx/d\phi)^2+(dy/d\phi)^2}}{\sqrt{2gy}}\,d\phi = \Bigl(\frac{R}{g}\Bigr)^{1/2}\phi.
\]
For a linear path, $x=L_l\cos\theta$, $y=L_l\sin\theta$, with $L_l=g\sin\theta\,t_l^2/2$:
\[
t_l=\Bigl(\frac{2L_l}{g\sin\theta}\Bigr)^{1/2}.
\]
The time ratio $\tau=t_{cyc}/t_l$ is:
\[
\tau=\left(\frac{(1-\cos\phi)\phi^2}{2(2+\phi^2-2\cos\phi-2\phi\sin\phi)}\right)^{1/2},
\]
and the distance ratio $\ell=\ell_l/\ell_{cyc}=\tau^2$. These are shown in
Fig.~\ref{Fig_Supp_inviscid_length_time_ratios}.

\begin{figure*}[!ht]
\centering{\includegraphics[width=.5\textwidth]{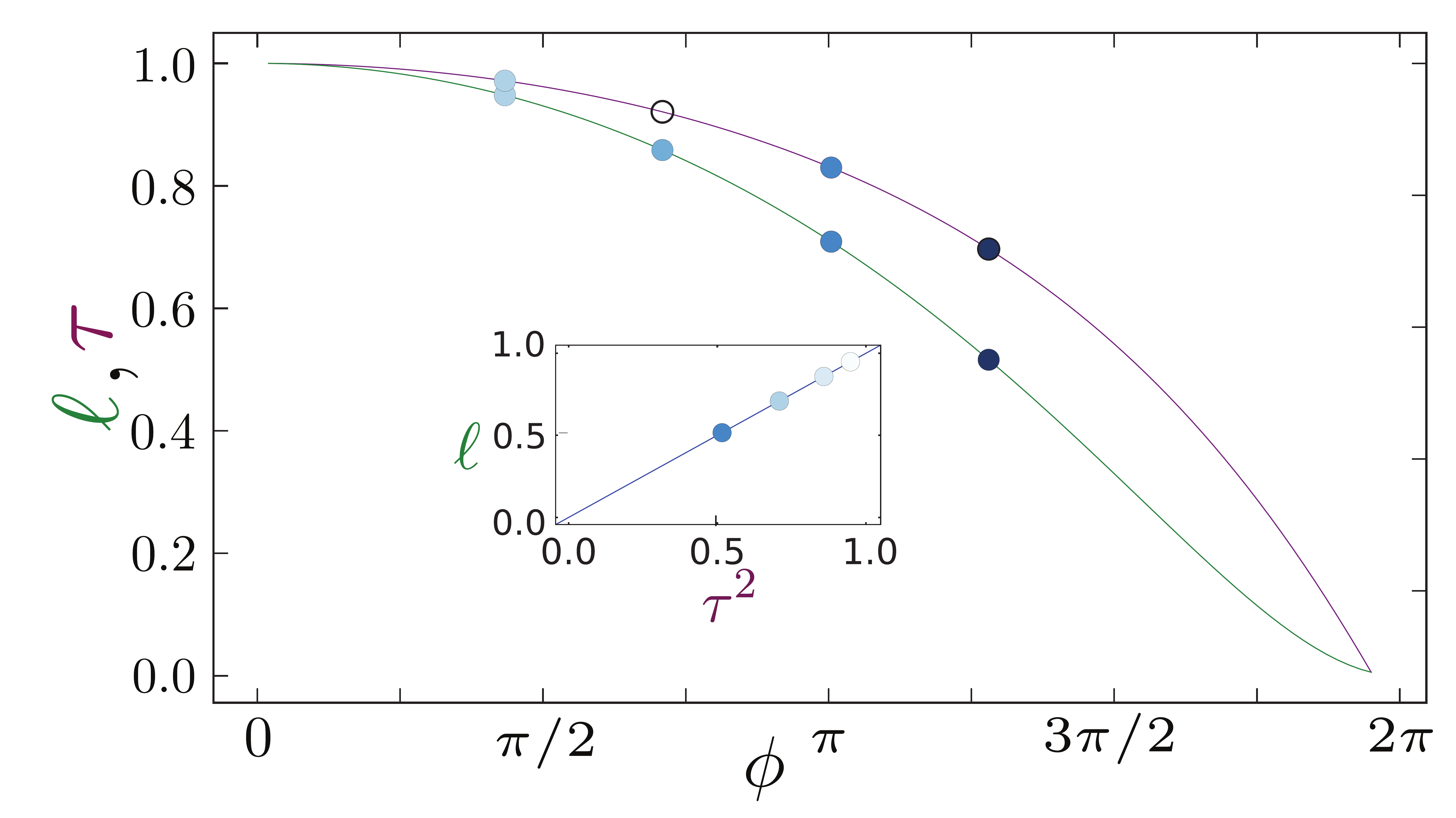}}
\caption{\label{Fig_Supp_inviscid_length_time_ratios}
Analytical ratios of travel time $\tau=t_{cyc}/t_l$ and path length $\ell=\ell_l/\ell_{cyc}$:
without fluid, the cycloid is both faster ($\tau<1$) and longer in distance ($\ell<1$).}
\end{figure*}

Accounting for buoyancy, inertia, rotational inertia, and a non-zero release
velocity $v_i$ gives a {\it vertically-shifted} cycloid $x=L(\phi-\sin\phi)$ and
$y=L(1-\cos\phi)+v_i^2(1+\tfrac{14}{5}\Gamma)/(2g(\Gamma-1))$.


\newpage 

\noindent
{\bf  Variational framework}

We seek the trajectory that minimizes travel time $\tilde{T}$ subject to a prescribed
energy-dissipation budget $\mathcal{E}_c$. By the standard isoperimetric method, the constrained
optimum coincides with the unconstrained minimum of the augmented functional
\begin{equation}
\mathcal{I}_\mu = \int_0^{\tilde{T}}\!\!\Bigl[1+\mu A\tilde{v}^2+\lambda(\tilde{v}'+A\tilde{v}-B\sin\theta)+\sigma_x(\tilde{x}'-\tilde{v}\cos\theta)+\sigma_y(\tilde{y}'-\tilde{v}\sin\theta)\Bigr]\,\mathrm{d}\tilde{t},
\tag{S-1}
\label{eq:Imu_SI}
\end{equation}
where $\mu\geq 0$ is a global constant enforcing the energy budget, and $\lambda$,
$\sigma_x$, $\sigma_y$ are pointwise multipliers. Setting $\mu=0$ recovers the pure time
functional; $\mu\to\infty$ recovers the energy-minimization functional. The admissible range is
$\mathcal{E}^{(E)}\leq\mathcal{E}_c<\mathcal{E}^{(T)}$.

First variation of $\mathcal{I}_\mu$ with respect to $\tilde{v}$, $\theta$, $\tilde{x}$,
$\tilde{y}$, $\lambda$, $\sigma_x$, $\sigma_y$:
\begin{subequations}
\begin{align}
\frac{\delta\mathcal{I}_\mu}{\delta\tilde{v}}=0\;&\Rightarrow\;
  2\mu A\tilde{v}+\lambda A-\sigma_x\cos\theta-\sigma_y\sin\theta-\lambda'=0,
  \tag{S-2a}\label{eq:S2a}\\
\frac{\delta\mathcal{I}_\mu}{\delta\theta}=0\;&\Rightarrow\;
  \sigma_x\tilde{v}\sin\theta-\lambda B\cos\theta-\sigma_y\tilde{v}\cos\theta=0,
  \tag{S-2b}\label{eq:S2b}\\
\frac{\delta\mathcal{I}_\mu}{\delta\tilde{x}}=0\;&\Rightarrow\;\sigma'_x=0,
  \tag{S-2c}\\
\frac{\delta\mathcal{I}_\mu}{\delta\tilde{y}}=0\;&\Rightarrow\;\sigma'_y=0,
  \tag{S-2d}\\
\frac{\delta\mathcal{I}_\mu}{\delta\lambda}=0\;&\Rightarrow\;\tilde{v}'+A\tilde{v}-B\sin\theta=0,
  \tag{S-2e}\label{eq:EOM}\\
\frac{\delta\mathcal{I}_\mu}{\delta\sigma_x}=0\;&\Rightarrow\;\tilde{x}'=\tilde{v}\cos\theta,
  \tag{S-2f}\\
\frac{\delta\mathcal{I}_\mu}{\delta\sigma_y}=0\;&\Rightarrow\;\tilde{y}'=\tilde{v}\sin\theta.
  \tag{S-2g}
\end{align}
\end{subequations}
Equations~(S-2c)--(S-2d) imply $\sigma_x$, $\sigma_y$ are constants. The free-endpoint
transversality condition ($\lambda_f=0$, variation with respect to $\tilde{T}$) gives
\begin{equation}
\sigma_x\tilde{v}_f\cos\theta_f+\sigma_y\tilde{v}_f\sin\theta_f = 1+\mu A\tilde{v}_f^2.
\tag{S-3}
\label{eq:transv_mu}
\end{equation}
Evaluating Eq.~(S-2b) at $\tilde{T}$ with $\lambda_f=0$ forces
$(\sigma_x,\sigma_y)\propto(\cos\theta_f,\sin\theta_f)$. Substituting into Eq.~(S-3):
\begin{equation}
\sigma_x=\frac{\Omega\cos\theta_f}{\tilde{v}_f},\quad
\sigma_y=\frac{\Omega\sin\theta_f}{\tilde{v}_f},\qquad
\Omega\equiv 1+\mu A\tilde{v}_f^2.
\tag{S-4}
\label{eq:sigmas_mu}
\end{equation}
At $\mu=0$: $\Omega=1$, recovering the time-min multipliers; as $\mu\to\infty$:
$\sigma_x/\sigma_y=\cos\theta_f/\sin\theta_f$, recovering the energy-min multipliers.

Since $\sigma_x$, $\sigma_y$ are constants, $\lambda=\tilde{v}(\sigma_x\tan\theta-\sigma_y)/B$,
giving $\lambda'=(\sigma_x\tan\theta-\sigma_y)\tilde{v}'/B+\sigma_x\tilde{v}\theta'/(B\cos^2\theta)$.
Substituting $\sigma_{x,y}$ from Eq.~(S-4) and $\lambda'$ into Eq.~(S-2a), using
$\tan\theta-\tan\theta_f=\sin(\theta-\theta_f)/(\cos\theta\cos\theta_f)$:
\begin{equation}
\frac{d\theta}{d\tilde{t}} =
\underbrace{-\frac{B\cos\theta}{\tilde{v}}+2A\cos^2\theta(\tan\theta-\tan\theta_f)}_{\text{time-min, Eq.~(S-11)}}
+\frac{2\Pi B\cos^2\theta}{\tilde{v}_f\cos\theta_f},
\tag{\textcolor{black}{S-5}}
\label{eq:dthetadt_mu_SI}
\end{equation}
where
\begin{equation}
\Pi\;\equiv\;\frac{\mu A\tilde{v}_f^2}{1+\mu A\tilde{v}_f^2}\;\in\;[0,1)
\tag{\textcolor{black}{S-6}}
\label{eq:Pi_SI}
\end{equation}
maps $\mu\in[0,\infty)$ onto $\Pi\in[0,1)$.
Dividing Eq.~(S-5) by $\tilde{v}$ gives the intrinsic curvature
$\kappa^{(\mu)}\equiv d\theta/ds$ (Eq.~3 of the main text):
\begin{equation}
\kappa^{(\mu)} =
\underbrace{-\frac{B\cos\theta}{\tilde{v}^2}+\frac{2A\cos\theta\sin(\theta-\theta_f)}{\tilde{v}\cos\theta_f}}_{\kappa^{(T)}}
+\Pi\cdot\underbrace{\frac{2B\cos^2\theta}{\tilde{v}\,\tilde{v}_f\cos\theta_f}}_{\kappa^{(E)}\,>\,0}.
\tag{\textcolor{black}{S-7}}
\label{eq:kappa_mu_SI}
\end{equation}
The unified ODE for $\tilde{v}(\theta)$, which governs the numerical computation of all
optimal paths and was used to generate Figs.~3 and 4 of the main text, follows from
Eqs.~(S-5) and (S-2e):
\begin{equation}
\frac{d\tilde{v}}{d\theta}=\frac{-A\tilde{v}+B\sin\theta}{\dfrac{2\Pi B\cos^2\theta}{\tilde{v}_f\cos\theta_f}-\dfrac{B\cos\theta}{\tilde{v}}+2A\cos^2\theta(\tan\theta-\tan\theta_f)}.
\tag{\textcolor{black}{S-8}}
\label{eq:dvdtheta_mu_SI}
\end{equation}
At $\Pi=0$ this is the time-min ODE; at $\Pi=1$ it is the energy-min ODE. The Beltrami
identity ($\mathcal{I}_\mu$ has no explicit time dependence) gives the constant of motion
\begin{equation}
(1+\mu A\tilde{v}^2)+\lambda(A\tilde{v}-B\sin\theta)-\tilde{v}(\sigma_x\cos\theta+\sigma_y\sin\theta)=0,
\tag{\textcolor{black}{S-9}}
\end{equation}
where $C_0=0$ follows from evaluating at $\tilde{T}$ using $\lambda_f=0$ and Eq.~(S-3).
The three unknown parameters $\theta_i$, $\theta_f$, $\tilde{v}_f$ are determined from
\begin{align}
\sqrt{1-\tilde{H}^2} &=\int_{\theta_i}^{\theta_f}\!\!\frac{\tilde{v}\cos\theta\;\mathrm{d}\theta}{\text{denom}},
\tag{\textcolor{black}{S-10a}}\\
\tilde{H} &=\int_{\theta_i}^{\theta_f}\!\!\frac{\tilde{v}\sin\theta\;\mathrm{d}\theta}{\text{denom}},
\tag{\textcolor{black}{S-10b}}\\
0 &=\frac{1+\mu A\tilde{v}_i^2}{\Omega}+\frac{\tilde{v}_i\sin(\theta_i-\theta_f)(A\tilde{v}_i-B\sin\theta_i)}{B\tilde{v}_f\cos\theta_f}-\frac{\tilde{v}_i\cos(\theta_i-\theta_f)}{\tilde{v}_f},
\tag{\textcolor{black}{S-10c}}
\end{align}
where denom $=2\Pi B\cos^2\theta/(\tilde{v}_f\cos\theta_f)-B\cos\theta/\tilde{v}+2A\cos^2\theta(\tan\theta-\tan\theta_f)$. For $\tilde{v}_i\to 0$, $\theta_i\to\pi/2$.

\vspace{0.5cm}
\noindent
{\bf Special case $\Pi=0$: time-minimizing paths}

Setting $\Pi=0$ ($\mu=0$, $\Omega=1$) in the unified framework: the multipliers reduce to
$\sigma_x=\cos\theta_f/\tilde{v}_f$, $\sigma_y=\sin\theta_f/\tilde{v}_f$. Equations~(S-5)
and~(S-8) become
\begin{equation}
\frac{d\theta}{d\tilde{t}} = -\frac{B\cos\theta}{\tilde{v}}+2A\cos^2\theta(\tan\theta-\tan\theta_f),
\tag{\textcolor{black}{S-11}}
\label{eq:dtimethetat}
\end{equation}
\begin{equation}
\frac{d\tilde{v}}{d\theta}= \frac{-A\tilde{v}+B\sin\theta}{-B\cos\theta/\tilde{v}+2A\cos^2\theta(\tan\theta-\tan\theta_f)},
\tag{\textcolor{black}{S-12}}
\label{eq:dvdtheta_time}
\end{equation}
and the endpoint conditions~(S-10a)--(S-10c) reduce to
\begin{equation}
\begin{aligned}
\sqrt{1-\tilde{H}^2} &= \int_{\theta_i}^{\theta_f}\frac{\tilde{v}\cos\theta\;\mathrm{d}\theta}{-B\cos\theta/\tilde{v}+2A\cos^2\theta(\tan\theta-\tan\theta_f)},\\
\tilde{H} &= \int_{\theta_i}^{\theta_f}\frac{\tilde{v}\sin\theta\;\mathrm{d}\theta}{-B\cos\theta/\tilde{v}+2A\cos^2\theta(\tan\theta-\tan\theta_f)},\\
\frac{\cos\theta_i}{\tilde{v}_i} &=\frac{\cos\theta_f}{\tilde{v}_f}+\frac{A\tilde{v}_i\sin\theta_f\cos\theta_i}{B\tilde{v}_f}-\frac{A\tilde{v}_i\cos\theta_f\sin\theta_i}{B\tilde{v}_f}.
\end{aligned}
\tag{\textcolor{black}{S-13}}
\label{eq:timefinalthetaS}
\end{equation}

\vspace{0.5cm}
\noindent
{\bf Intrinsic curvature of the time-minimizing path:}
This is the $\Pi=0$ limit of Eq.~(S-7). Dividing Eq.~(S-11) by $\tilde{v}$:
\begin{equation}
\kappa^{(T)} = -\frac{B\cos\theta}{\tilde{v}^2} +
\frac{2A\cos\theta}{\tilde{v}}\cdot\frac{\sin(\theta-\theta_f)}{\cos\theta_f}.
\tag{\textcolor{black}{S-14}}
\label{eq:kappa_time_SI}
\end{equation}
The first term is the inertia-dominated contribution, identical to the
curvature of a classical cycloid ($A\to 0$); the second is the viscous correction, vanishing at
$\theta=\theta_f$ and largest in the midsection.

The two-term structure of Eq.~(S-14) explains the composite character of the
optimal fastest path discussed in the main text: near the endpoints the particle is slow and the
inertial term $-B\cos\theta/\tilde{v}^2$ dominates, producing cycloidal end-caps; in the
mid-section the particle approaches terminal velocity and the viscous correction straightens
the path into a near-linear ramp.

\noindent\textit{Sign of endpoint curvature.} At $\theta=\theta_f$ the viscous correction vanishes:
\begin{equation*}
\kappa^{(T)}_f = -\frac{B\cos\theta_f}{\tilde{v}_f^2} < 0,
\end{equation*}
confirming the concave ($\kappa<0$, bowl-shaped) end-cap for any $St_p$.

\noindent\textit{End-cap scaling in the viscous limit.} For $St_p\ll 1$:
$\tilde{v}_f\approx(B/A)\sin\theta_f$, $\sin\theta_f=\tilde{H}$:
\begin{equation*}
\frac{1}{|\kappa^{(T)}_f|} = \frac{\tilde{v}_f^2}{B\cos\theta_f}
\xrightarrow{St_p\ll 1}
\frac{B\sin^2\theta_f}{A^2\cos\theta_f} = 2\tan\theta_f\cdot St_p^2\tilde{H}^2
\equiv 2\tan\theta_f\cdot L_{char},
\end{equation*}
where $\tan\theta_f=\tilde{H}/\sqrt{1-\tilde{H}^2}$ is fixed entirely by
endpoint geometry. This prediction is confirmed in Fig.~3b of the main text.

\vspace{0.5cm}
\noindent
{\bf Numerical perturbation tests (time-minimizing paths):} Multiple perturbation trajectories
(\#1 to \#14) are compared in Fig.~\ref{Fig_time_perturbation_viscid}a, with curves $\#6$, $\#7$,
$\#8$ as the theoretical minimum time trajectories for three distinct Stokes numbers. In all
panels, every perturbation curve yields time strictly greater than the minimum, validating the
robustness of the computed optimal trajectories.

\begin{figure*}[!ht]
\centering{\includegraphics[width=1\textwidth]{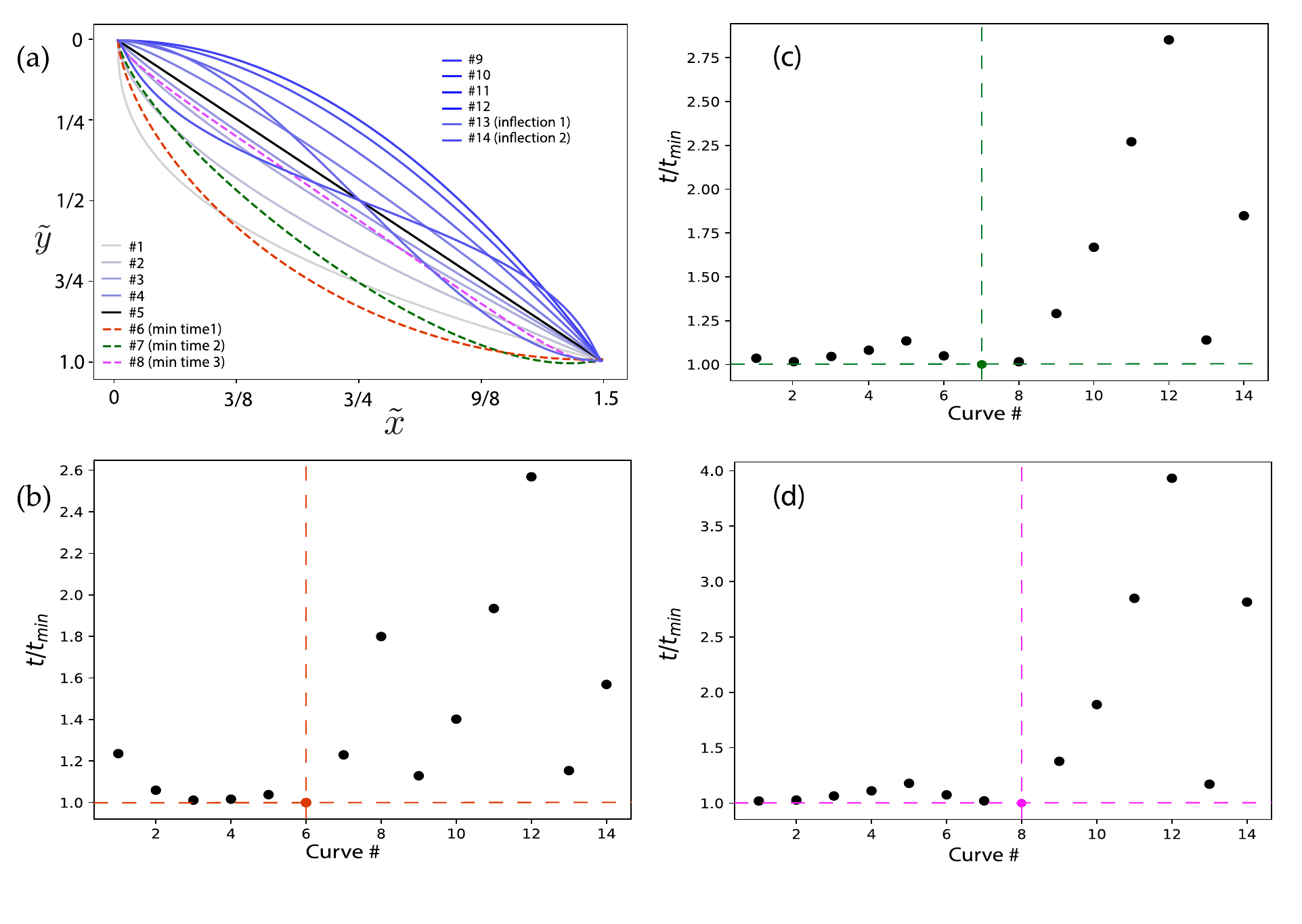}}
\caption{\label{Fig_time_perturbation_viscid}
Numerical tests of minimum time on perturbations of the optimal paths. {(a) Families of trial trajectories; in each family the minimum-time path is one of curves $\#6$, $\#7$, $\#8$, corresponding to three different Stokes numbers $St_p$.} All
variables are non-dimensional ($\tilde{x}$, $\tilde{y}$, $\tilde{t}$). {Curves $\#6$--$8$ are the true minimum-time paths for three different $St_p$.} {(b)--(d) Non-dimensional time ratio $\tilde{t}/\tilde{t}_{min}$ for each of the three Stokes numbers in (a):}{(b)--(d) Non-dimensional time ratio $\tilde{t}/\tilde{t}_{min}$:} every perturbation yields strictly $\tilde{t}>\tilde{t}_{min}$,
validating the optimality of the computed trajectories.}
\end{figure*}

\vspace{0.5cm}
\noindent
{\bf Special case $\Pi\to 1$: energy-minimizing paths}

Setting $\Pi\to 1$ ($\mu\to\infty$) in the unified framework: the multipliers become
$\sigma_x=A\tilde{v}_f\cos\theta_f$, $\sigma_y=A\tilde{v}_f\sin\theta_f$. Equations~(S-5)
and~(S-8) become
\begin{equation}
\frac{d\theta}{d\tilde{t}} = \frac{2B\cos^2\theta}{\tilde{v}_f\cos\theta_f}-\frac{B\cos\theta}{\tilde{v}}+2A\cos^2\theta(\tan\theta-\tan\theta_f),
\tag{\textcolor{black}{S-15}}
\label{eq:dthetadt_energy}
\end{equation}
\begin{equation}
\frac{d\tilde{v}}{d\theta} = \frac{-A\tilde{v}+B\sin\theta}{\dfrac{2B\cos^2\theta}{\tilde{v}_f\cos\theta_f}-\dfrac{B\cos\theta}{\tilde{v}}+2A\cos^2\theta(\tan\theta-\tan\theta_f)},
\tag{\textcolor{black}{S-16}}
\label{eq:dvdtheta_energy}
\end{equation}
and the endpoint conditions~(S-10a)--(S-10c) reduce to
\begin{equation}
\begin{aligned}
\sqrt{1-\tilde{H}^2} &= \int_{\theta_i}^{\theta_f}\frac{\tilde{v}\cos\theta\;\mathrm{d}\theta}{\dfrac{2B\cos^2\theta}{\tilde{v}_f\cos\theta_f}-\dfrac{B\cos\theta}{\tilde{v}}+2A\cos^2\theta(\tan\theta-\tan\theta_f)},\\
\tilde{H} &= \int_{\theta_i}^{\theta_f}\frac{\tilde{v}\sin\theta\;\mathrm{d}\theta}{\dfrac{2B\cos^2\theta}{\tilde{v}_f\cos\theta_f}-\dfrac{B\cos\theta}{\tilde{v}}+2A\cos^2\theta(\tan\theta-\tan\theta_f)},\\
\frac{\tilde{v}_i\cos\theta_i}{\cos\theta_f} &= \tilde{v}_f\!\left[1+\frac{A\tilde{v}_i\cos\theta_i}{B}(\tan\theta_f-\tan\theta_i)\right].
\end{aligned}
\tag{\textcolor{black}{S-17}}
\label{eq:energyfinalthetaS}
\end{equation}
The condition $\tilde{v}_i=0$ necessitates $\theta_f=\pi/2$ (free vertical fall), which is
not a valid solution for arbitrary $(x_f,y_f)$.

\vspace{0.5cm}
\noindent
{\bf Intrinsic curvature of the energy-minimizing path:}
This is the $\Pi\to 1$ limit of Eq.~(S-7). Dividing Eq.~(S-15) by $\tilde{v}$:
\begin{equation}
\kappa^{(E)} = \frac{2B\cos^2\theta}{\tilde{v}\,\tilde{v}_f\cos\theta_f}
- \frac{B\cos\theta}{\tilde{v}^2}
+ \frac{2A\cos\theta}{\tilde{v}}\cdot\frac{\sin(\theta-\theta_f)}{\cos\theta_f}.
\tag{\textcolor{black}{S-18}}
\label{eq:kappa_energy_SI}
\end{equation}
The structural difference from the time-min case is the $\kappa^{(E)}$ contribution
(third term in Eq.~S-7, switched on fully at $\Pi=1$):
\begin{equation*}
\kappa^{(E)}_{\text{full}}-\kappa^{(T)} =
\frac{2B\cos^2\theta}{\tilde{v}\,\tilde{v}_f\cos\theta_f} > 0
\quad\text{everywhere, for any }St_p.
\end{equation*}

Since this difference is positive at every point along the path, the energy-minimizing curve is
strictly more convex than the fastest path. Physically, the positive $\kappa^{(E)}$ term pushes
the trajectory toward shallower slopes, keeping the velocity low and thereby reducing the
cumulative viscous dissipation $\int A\tilde{v}^2\,d\tilde{t}$ at the cost of longer travel
time.

\noindent\textit{Sign of endpoint curvature.} At $\theta=\theta_f$,
$\sin(\theta_f-\theta_f)=0$:
\begin{equation*}
\kappa^{(E)}_f = \frac{2B\cos\theta_f}{\tilde{v}_f^2}-\frac{B\cos\theta_f}{\tilde{v}_f^2}
= +\frac{B\cos\theta_f}{\tilde{v}_f^2} > 0,
\end{equation*}
while $\kappa^{(T)}_f=-B\cos\theta_f/\tilde{v}_f^2<0$. The sign reversal is
algebraically exact and valid for any $St_p$: the $\kappa^{(E)}$ term, evaluated at
$\theta_f$, doubles the inertial contribution and reverses its sign.

\noindent\textit{Curvature scaling in the viscous limit.} For $St_p\ll 1$:
$\tilde{v}_f\to(B/A)\sin\theta_f$:
\begin{equation*}
\frac{1}{|\kappa^{(E)}_f|} = \frac{\tilde{v}_f^2}{B\cos\theta_f}
\xrightarrow{St_p\ll 1}
\frac{B\sin^2\theta_f}{A^2\cos\theta_f} = 2\tan\theta_f\cdot St_p^2\tilde{H}^2 \equiv
2\tan\theta_f\cdot L_{char},
\end{equation*}
identical in magnitude to the time-min scaling but with opposite sign:
a precise, parameter-free curvature duality.

\vspace{0.5cm}
\noindent
{\bf Limit of least dissipation:} For $St_p\gg 1$ ($A\approx 0$), integrating Eq.~(S-16) gives
$v=v_i\cos\theta_i/\cos\theta$ (taking the $c_1=0$, $+$ branch). Substituting into Eq.~(S-17)
and integrating:
\begin{align}
x &= \frac{v^2_i(\frac{7}{5}\Gamma+\frac{1}{2})\cos^2\theta_i}{(\Gamma-1)g}(\tan\theta-\tan\theta_i), \notag\\
y &= \frac{v^2_i(\frac{7}{5}\Gamma+\frac{1}{2})}{2(\Gamma-1)g\cos^2\theta}(\cos^2\theta_i-\cos^2\theta).
\tag{\textcolor{black}{S-19}}
\end{align}
Eliminating $\theta$ yields the parabola
\begin{equation}
y=\frac{B(1+\tan^2\theta_i)}{2v_i^2}x^2-x\tan\theta_i,
\tag{\textcolor{black}{S-20}}
\end{equation}
which coincides with the equations of projectile motion.

These parabolic trajectories are shown in Fig.~\ref{Fig_parabola_supp} for several values
of the initial velocity $\tilde{v}_i$. For a given $\tilde{v}_i$, two distinct parabolas exist;
the energy-minimizing path is the shorter of the two. {Numerical perturbation tests confirm that these least-dissipation parabolas are true energy minima (Fig.~\ref{Fig_energy_perturbation_inviscid}).}

\begin{figure*}[!ht]
\centering{\includegraphics[width=0.5\textwidth]{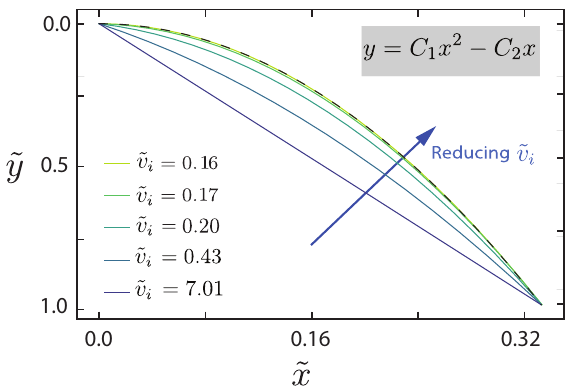}}
\caption{\label{Fig_parabola_supp}
Energetic minima parabolas corresponding to the {\it least dissipation limit}
($St_p\gg 1$, $A\to 0$). Each curve is the energy-minimizing trajectory (Eq.~S-20) for
a different initial velocity $\tilde{v}_i$. The parabolic form $y = C_1 x^2 - C_2 x$ coincides
exactly with Newton's projectile equations, confirming the recovery of classical ballistic
motion in the absence of viscous dissipation.}
\end{figure*}

\begin{figure*}[!ht]
\centering{\includegraphics[width=1\textwidth]{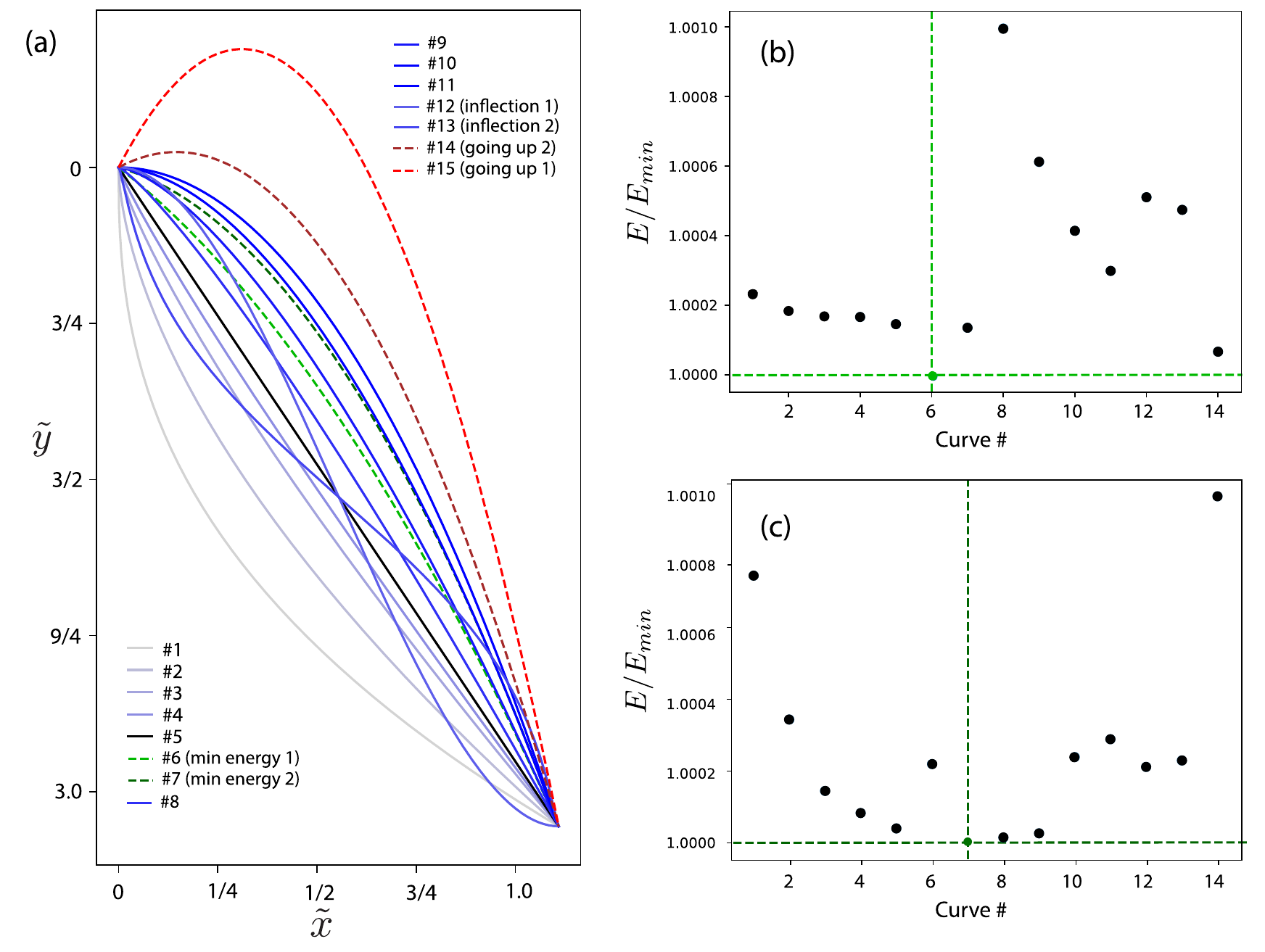}}
\caption{\label{Fig_energy_perturbation_inviscid}
Numerical tests of minimum energy on perturbations (least dissipation limit). All variables are
non-dimensional ($\tilde{x}$, $\tilde{y}$, $\tilde{t}$). Curves $\#6$--$7$ are the true minimum
energy paths for two different initial velocities $\tilde{v}_i$; $\#14$--$\#15$ are the
corresponding ascending parabolas. (b)--(c) Energy ratio $\tilde{E}/\tilde{E}_{min}$: all
perturbations yield higher energy than the minimum, confirming optimality.}
\end{figure*}

\vspace{0.5cm}
\noindent
{\bf Numerical perturbation tests (energy-minimizing paths, viscous regime):} Multiple
perturbation trajectories (\#1 to \#14) are compared in Fig.~\ref{Fig_energy_perturbation_all}a,
with curves $\#6$, $\#7$, $\#8$ as the theoretical minimum energy trajectories for three distinct
Stokes numbers. In all panels, every perturbation curve yields energy strictly greater than the
minimum.

\begin{figure*}[!ht]
\centering{\includegraphics[width=1\textwidth]{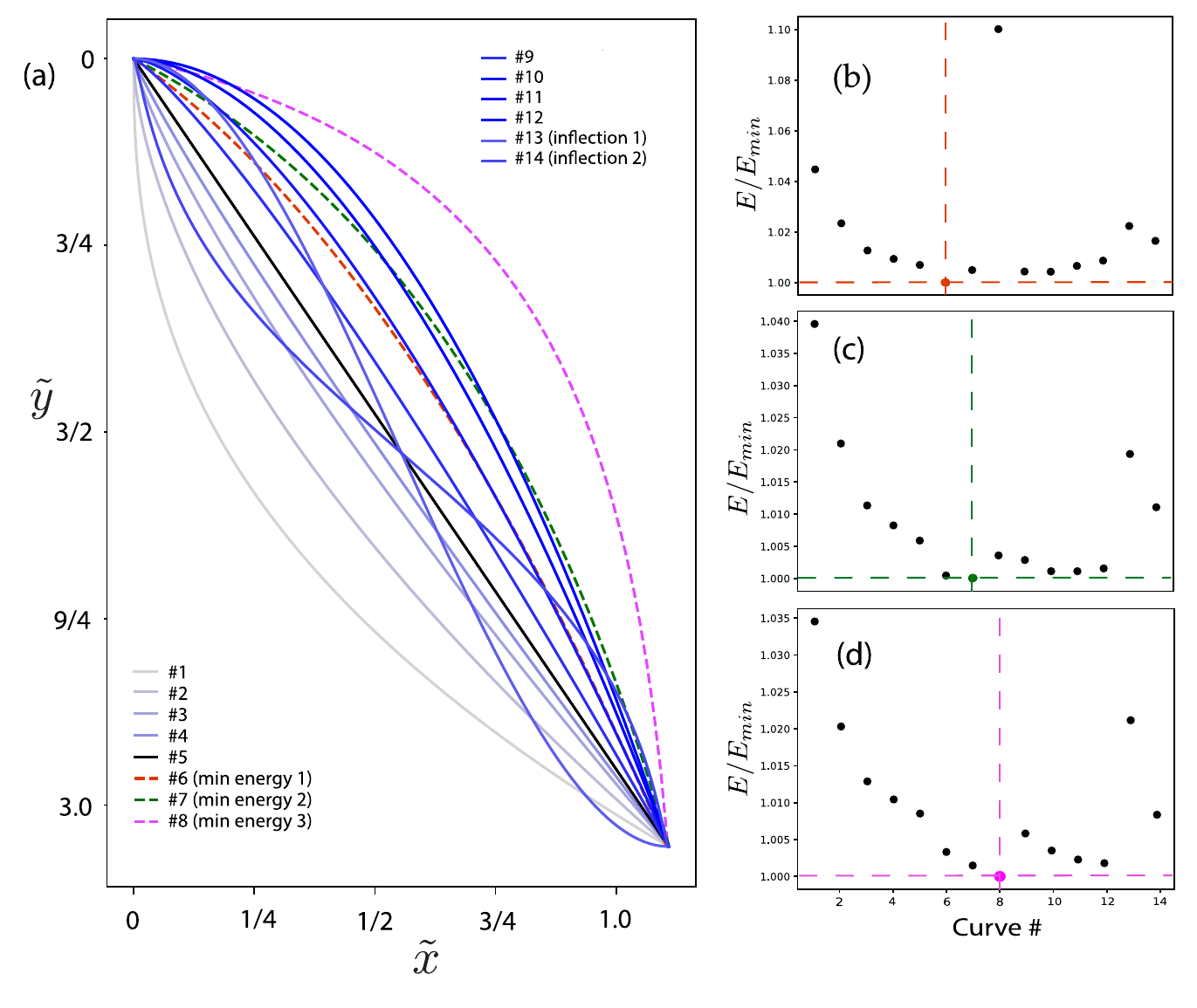}}
\caption{\label{Fig_energy_perturbation_all}
Numerical tests of minimum energy on perturbations (viscous regime). Curves $\#6$--$8$ are the
true minimum energy curves for three different Stokes numbers. (b)--(d) Energy ratio $E/E_{min}$:
all perturbations yield higher energy than $E_{min}$.}
\end{figure*}

\vspace{0.5cm}
\noindent
{\bf General $\Pi$: conflicting starting and ending curvatures}

The unified curvature equation~(S-7) holds for all $\Pi\in[0,1)$. We analyze its
behavior at both ends of every path.

\noindent\textit{Near the start.} As $\theta\to\pi/2$ and $\tilde{v}\to 0$: both
$\cos\theta\sim\epsilon$ and $\tilde{v}\sim c\epsilon$ (from the EOM at low speed), so
$\cos\theta/\tilde{v}^2\sim 1/\epsilon\to+\infty$ and the inertial term diverges to $-\infty$.
The viscous correction stays $\mathcal{O}(1)$, and the $\Pi$-correction $\propto\cos^2\theta/\tilde{v}\sim\epsilon\to 0$ vanishes regardless of $\Pi$. Hence
\begin{equation*}
\kappa^{(\mu)}\xrightarrow{\tilde{v}\to 0}-\infty\qquad\text{for all }\Pi\in[0,1).
\end{equation*}
\noindent\textit{At the endpoint.} At $\theta=\theta_f$, $\sin(\theta_f-\theta_f)=0$, so
the viscous correction vanishes and Eq.~(S-7) reduces exactly to
\begin{equation*}
\kappa_f^{(\mu)} = \frac{B\cos\theta_f}{\tilde{v}_f^2}(2\Pi-1):
\end{equation*}
negative for $\Pi<1/2$, zero at $\Pi^*=1/2$, positive for $\Pi>1/2$.

\noindent\textit{S-shaped trajectories.} Since $\kappa^{(\mu)}\to-\infty$ at the start while
$\kappa_f^{(\mu)}>0$ for $\Pi>1/2$, the intermediate value theorem guarantees at least one zero
crossing (an inflection point) in the interior of the path. The constrained-optimal path is
therefore S-shaped for all $\Pi>1/2$.
In the inviscid limit ($A\to 0$), the pure energy minimum ($\Pi\to 1$) is itself a convex
parabola without inflection (Eq.~S-20); the S-shape is therefore a distinctive signature of the
constrained optimization, rather than of
the pure energy minimum.
The inflection sits exactly at the endpoint at $\Pi^*=1/2$ and migrates toward the
interior as $\Pi$ increases beyond $1/2$. For $\Pi<1/2$, both endpoint values of $\kappa$ are
negative and no inflection is guaranteed.

\end{adjustwidth}
\endgroup
\twocolumngrid

\end{document}